\documentclass[12pt,a4paper]{article}
\usepackage{jheppub}  
\usepackage{amssymb} 
\usepackage{amsmath}
\usepackage{mathtools}
\usepackage{amsfonts}    
\usepackage{dsfont}
\usepackage{pdfpages}
\usepackage{verbatim}
\usepackage{tensor}
\usepackage{mathrsfs}
\usepackage[mathscr]{euscript}
\usepackage{tikz}
\usetikzlibrary{patterns,decorations.pathreplacing}
\usepackage{slashed}

\newcommand{\ba}{\begin{align}}

\newcommand{\be}{\begin{equation}}
\newcommand{\ee}{\end{equation}}
\def\bd{\begin{tikzpicture}}
\def\ed{\end{tikzpicture}}

\renewcommand\Im{\mathop{\text{Im}}}

\renewcommand\Re{\mathop{\text{Re}}}

\usepackage{tcolorbox}
  
\allowdisplaybreaks[1]

\newcommand\vol{\mathop{\text{vol}}}
\title{The Disk Partition Function in String Theory}

\author{Lorenz Eberhardt\,,} 
\author{Sridip Pal} 
\affiliation{School of Natural Sciences, Institute for Advanced Study,\\ 1 Einstein Drive,
Princeton,  NJ 08540, USA}
\emailAdd{elorenz@ias.edu}
\emailAdd{sridip@ias.edu}

\abstract{
We investigate the disk partition function for the open string. This is a subtle problem because of the presence of a residual gauge group $\mathrm{PSL}(2,\mathbb{R})$ on the worldsheet even after fixing the conformal gauge. It naively has infinite volume and leads to a vanishing answer. We use different methods that all demonstrate that $\mathrm{PSL}(2,\mathbb{R})$ effectively behaves like a group with finite negative volume in the path integral, which leads to a simple prescription for the computation of the disk partition function. We apply our findings to give a simple rederivation of the D-brane tensions.

}

\begin{document}

\maketitle

%make math in all titles bold
\makeatletter
\g@addto@macro\bfseries{\boldmath}
\makeatother
%end code

%%%%%%%%%%%%%%%%%%%%%%%%%%%%%%%%%%%%%%%%%%%%%%%%%%%%%%%%%%%%%%%
\section{Introduction}
In string perturbation theory much effort was devoted historically to understand higher point and higher genus correlation functions. For a broad overview, see e.g.~\cite{DHoker:1988pdl, Witten:2012bh}. Despite a good understanding of the integrands of string perturbation theory, performing the actual integrals has remained a challenging task. 

On the other end of the spectrum, there are some exceptional correlators at genus 0 that require special attention. The reason for this is a residual gauge group after imposing conformal gauge which is present due to conformal Killing vectors. For the sphere, there are three complex conformal Killing vectors corresponding to the group of M\"obius transformations. Since the volume of this group is infinite, one naively concludes that zero-point, one-point and two-point functions vanish at tree-level in string theory. The same goes for the open string, where the group of residual M\"obius transformations in $\mathrm{PSL}(2,\mathbb{R})$. This conclusion is however premature, since the infinities of the residual gauge groups can potentially be compensated by other infinities in the worldsheet path integral. It is a subtle problem to compute the actual value of these quantities and only a partial understanding exists, see \cite{Tseytlin:1987ww, Liu:1987nz, Tseytlin:1988tv, Erbin:2019uiz}. Various such quantities were also successfully computed for strings on $\text{AdS}_3$ \cite{Maldacena:2001km, Troost:2011ud}.

All these quantities have a physical meaning on which we would like to comment. Zero-point functions represent the on-shell value of the action of the effective spacetime theory, which is (super)gravity in the case of the closed string and the D-brane worldvolume gauge theory in the case of the open string. These quantities are generically non-vanishing and especially in the case of the gravity on-shell action somewhat subtle to define. To get a finite answer one has to introduce local counterterms on an asymptotic cutoff surface. The first of these is the Gibbons-Hawking-York boundary term \cite{Gibbons:1976ue}. Introducing a cutoff in spacetime would be inconsistent with Weyl symmetry in string theory and it is unclear in general how to implement it in string theory. We consider this a very important open problem in understanding the emergence of gravity from string theory. 

One-point functions for the closed string represent tadpole diagrams in spacetime. Most of these tadpole diagrams vanish due to the spacetime equations of motion. There are however interesting non-vanishing one-point functions in string theory such as the dilaton one-point function or the example considered in \cite{Troost:2011ud}. 

Two-point functions represent the tree-level propagators of the spacetime theory. It was explained in \cite{Erbin:2019uiz} that these two-point functions are actually non-zero because the momentum conserving $\delta$-function $\delta^D(k_1-k_2)$ in spacetime is divergent thanks to the mass-shell condition that implies the conservation of the last component of the momenta provided that the other components are conserved. The correct expression in flat space is instead $2k^0 (2\pi)^{D-1} \delta^{D-1}(\vec{k}'-\vec{k})$. 
\medskip

In this paper, we give a reasonably complete understanding of the disk partition function, i.e.\ the open string zero-point function. The disk partition function computes interesting quantities directly in string theory such as D-brane tensions. Historically they are often computed in a roundabout way by imposing various consistency conditions for the exchange of closed strings between two parallel D-branes. The challenge in this computation is the presence of the residual gauge group $\mathrm{PSL}(2,\mathbb{R})$. Since this group is non-compact, it naively has infinite volume. However, it was proposed in \cite{Liu:1987nz} that it essentially behaves as a group with finite \emph{negative} volume in any computation so that the string disk partition function $Z_\text{disk}$ is simply related to worldsheet disk partition function $Z_\text{CFT}$ by
\be 
Z_\text{disk}=\frac{Z_\text{CFT}}{\vol(\mathrm{PSL}(2,\mathbb{R}))}\ .
\ee
This volume can be defined by a procedure akin to defining the gravitational on-shell action. In the normalization where the Ricci scalar on the group on the group with respect to the biinvariant metric is $\mathcal{R}=-6$, this volume works out to be $-\frac{\pi^2}{2}$. It is however very mysterious (at least to the authors) why this procedure should give the correct result.

We are thus motivated to reconsider the problem. We give in this paper three rigorous (for physicists' standards) ways to compute the disk partition function from first principles. Each of the methods reproduce this value for the effective volume. The first two methods are based on fixing a further gauge beyond the conformal gauge. Since the metric is already completely fixed, the further gauge fixing will invariably involve the matter fields on the worldsheet. For this reason we assume that the spacetime theory on which the string is propagating involves at least one flat direction, i.e.\ is for example time-independent. Backgrounds such as $\mathrm{AdS}_3 \times \mathrm{S}^3 \times  \mathbb{T}^4$ also work, since the torus directions are flat. We think however that our method can be generalized to other backgrounds as well. We explore two different gauge fixing conditions in terms of the free boson $X$ describing the flat target space direction. Both of them are slightly subtle and we discuss them in detail. One can gauge fix the worldsheet path integral further and compute the effective volume of the gauge group directly in this way. 
In the third method, we compute the disk partition function by relating it to a one-point function on the disk which can be computed without problems. This is done by assuming that the flat direction is compact. This introduces a modulus in the problem and the derivative of the disk partition function with respect to the modulus is by conformal perturbation theory given by a one-point function. We again recover the effective volume of $\mathrm{PSL}(2,\mathbb{R})$.

We finally apply this technique of computing disk partition functions to a short rederivation of D-brane tensions \cite{Polchinski:1995mt}. Since all relevant issues already arise for the bosonic string, we restrict to it for technical simplicity. We mention some open problems in Section~\ref{sec:conclusions}.

\section{\texorpdfstring{Gauge fixing $\boldsymbol{X_{\ell,m}=0}$}{Gauge fixing Xl,m=0}} \label{sec:first gauge}
We fix conformal gauge on the disk. In this section, it is convenient to use the upper hemisphere metric on the disk:
\be \label{eq:metric}
\hat{g}=\frac{4 \, \mathrm{d}z \, \mathrm{d}\bar{z}}{(1+|z|^2)^2}\ , \qquad |z | \le 1\ .
\ee
Any physical result will of course be independent of this choice because the full worldsheet theory is Weyl-invariant.
This form of the metric is convenient, because there is a standard orthonormal basis for the space of $L^2$-functions given by the spherical harmonics. We can consider two function spaces given by $L^2_\text{D}(D)$ and $L^2_\text{N}(D)$, where $D$ denotes here and in the following the disk. The former consists of all square-integrable functions $f$ on the unit disk satisfying Dirichlet boundary conditions $f(|z|=1)=0$,\footnote{We could generalize this to $f(|z|=1)=x_0$ for some constant $x_0$, but this constant could be removed by a spacetime translation.} while the latter consist of all square-integrable functions satisfying Neumann boundary conditions $\partial_n f(|z|=1)=0$, where $\partial_n$ is the normal (radial) derivative.

Spherical harmonics are given by $Y_{\ell,m}$, $\ell=0$, $1$, $2$, $\dots$ and $m=-\ell$, $-\ell+1$, $\dots$, $\ell$. They satisfy Neumann (Dirichlet) boundary conditions for $\ell+m \in 2\mathbb{Z}$ ($\ell+m \in 2\mathbb{Z}+1$). As we mentioned in the Introduction, we assume that there is one flat direction in spacetime which is described by the worldsheet boson $X$. In the following we will concentrate our attention on this boson. 
We can expand it into spherical harmonics
\be 
X=\sum_{\ell,m} X_{\ell,m} Y_{\ell,m} \label{eq:spherical harmonics expansion}
\ee
with $X_{\ell,m}=0$ for $\ell+m \in 2\mathbb{Z}+1$ and Neumann boundary conditions or $\ell+m \in 2\mathbb{Z}$ and Dirichlet boundary conditions.  Moreover, reality of $X$ imposes $X_{\ell,m}=\overline{X_{\ell,-m}}$.

Even after fixing the conformal gauge, there is a remaining gauge freedom that is not fully fixed. This is given by the group of conformal transformations, which acts as
\be 
X(z) \longmapsto X \circ \gamma^{-1}(z)
\ee
on the free boson $X$
 and fixes $g$. The latter is achieved by combining the diffeomorphism $\gamma$ with an appropriate Weyl transformation. The (global) conformal group on the disk is $\mathrm{PSU}(1,1) \cong \mathrm{PSL}(2,\mathbb{R})$ and acts by fractional linear transformations.\footnote{$\mathrm{PSL}(2,\mathbb{R})$ naturally acts on the upper half plane, whereas $\mathrm{PSU}(1,1)$ naturally acts on the unit disk. The two groups are isomorphic via the Cayley transform. We mostly use the name $\mathrm{PSL}(2,\mathbb{R})$.} Thus we have a path integral schematically of the following form
\be 
Z_{\text{disk}}=\int\frac{\mathscr{D}X}{\mathop{\text{vol}}(\mathrm{PSL}(2,\mathbb{R}))} \ \mathrm{e}^{-S[X]}\ .
\ee
The path integral over the appropriate space of functions (either $L^2_\text{N}(D)$ or $L^2_\text{D}(D)$). We remark that we have suppressed the presence of the ghosts and the other bosons in the path integral. Only with their presence the conformal anomaly cancels and it makes sense to gauge $\mathrm{PSL}(2,\mathbb{R})$.

Liu and Polchinski \cite{Liu:1987nz} provided with a prescription to calculate the ``regularized'' finite volume of the the group $\mathrm{PSL}(2,\mathbb{R})$, which we review in Appendix~\ref{app:volume PSL2R}. Using that, one can obtain 
\begin{equation}
Z_{\text{disk}}=-\frac{2}{\pi^2} \int \mathscr{D}X \ \mathrm{e}^{-S[X]}\ .
\end{equation}
Here one tacitly assumes a particular normalization of the ghost zero modes. This issue is also discussed in Appendix~\ref{app:volume PSL2R}.
We denote the CFT path integral that appear on the RHS by $Z_\text{CFT}$, 
\begin{equation}
Z_{\text{CFT}}\equiv \int \mathscr{D}X \ \mathrm{e}^{-S[X]}\ .
\end{equation}
We emphasize that the calculation of $Z_{\text{CFT}}$ does not gauge the global conformal group $\mathrm{PSL}(2,\mathbb{R})$.  

In what follows, we are going to show that
\be\label{eq:Main}
\frac{Z_{\text{disk}}}{Z_{\text{CFT}}}=-\frac{2}{\pi^2}
\ee
using standard QFT techniques, rather than calculating the regularized volume of $\mathrm{PSL}(2,\mathbb{R})$. Thus we want to also fix gauge-fix the global conformal group $\mathrm{PSL}(2,\mathbb{R})$. We achieve this by a slightly modified Faddeev-Popov procedure.

\subsection{Gauge choice and admissibility} \label{subsec:gauge choice}
The group of M\"{o}bius transformations preserving the unit disk is
\be 
\mathrm{PSU}(1,1)=\left\{ \begin{pmatrix}
a & b \\ \bar{b} & \bar{a} 
\end{pmatrix} \, \Big| \, |a|^2-|b|^2=1 \right\}\Big/ \sim\ .
\ee
Here, the equivalence $\sim$ identifies the matrix with the negative matrix. 
Only the $\mathrm{U}(1)$ subgroup specified by $b=0$ acts by isometries on the metric.  This realization of $\mathrm{PSU}(1,1)$ leads to a natural normalization of the biinvariant metric that is induced from ambient $\mathbb{C}^2 \cong \mathbb{R}^4$. This is the normalization which we shall use in the following. The explicit measure is given in Appendix~\ref{app:volume PSL2R}.

We would like to impose the gauge
\be 
X_{\ell,\pm m}=0
\ee
for some choice of $(\ell,m)$ in the expansion eq.~\!\eqref{eq:spherical harmonics expansion}. Note that due to the reality condition $\overline{X_{\ell,m}}=X_{\ell,-m}$, this is one complex or two real conditions. 
This fixes all non-compact directions of $\mathrm{PSL}(2,\mathbb{R}) \cong \mathrm{PSU}(1,1)$ and only leaves the Cartan subgroup $\mathrm{U}(1)$ unbroken. Since its volume is finite it is easy to take this into account. For concreteness, let us consider the following two gauge fixing conditions:
\begin{tcolorbox}
\be 
\text{Dirichlet:}\ X_{2,\pm 1}=0\ , \qquad \text{Neumann:}\ X_{1,\pm 1}=0\ . \label{Gauge}
\ee
\end{tcolorbox}
In what follows we will be proving the admissibility of the gauge choice. The argument for $m\not \in  \{-1,1\}$ is analogous and will lead to the same final result.

\paragraph{Admissibility of gauge choice.}
Since the Cartan generator $\mathrm{U}(1) \subset \mathrm{PSU}(1,1)$ remains unbroken, it is convenient to consider the coset $\mathrm{PSU}(1,1)/\mathrm{U}(1) \cong D$, which can also be identified with the unit disk. We stress that this unit disk is not the worldsheet! It comes equipped with a hyperbolic metric that descends from $\mathrm{PSU}(1,1)$, which takes the form for $\alpha \in D$
\be 
g=\frac{\pi \, \mathrm{d} \alpha \, \mathrm{d} \bar{\alpha}}{(1-|\alpha|^2)^2}\ .
\ee
The normalization is induced from the Haar measure on $\mathrm{PSU}(1,1)$. An explicit representative of $\alpha$ in $\mathrm{PSU}(1,1)$ is given by
\be 
\gamma_\alpha=\frac{1}{\sqrt{1-|\alpha|^2}}
\begin{pmatrix}
1 & \alpha \\
\bar{\alpha} & 1
\end{pmatrix}\ . \label{eq:gammaalpha}
\ee
This M\"{o}bius transformation has the property that $\gamma_\alpha(0)=\alpha$. 
To be explicit, the gauge conditions in eq.~\eqref{Gauge} read respectively
\begin{subequations}
\begin{align}\label{eq:Dg}
\text{Dirichlet}&: \qquad\int_{D}\frac{4\,\mathrm{d}^2 z}{(1+|z|^2)^2} X \circ \gamma_\alpha^{-1} (z,\bar{z}) Y_{2,1}(\bar z, z)
=0\ ,\\
\text{Neumann}&:\qquad
\int_{D}\frac{4\,\mathrm{d}^2 z}{(1+|z|^2)^2} X \circ \gamma_\alpha^{-1} (z,\bar{z}) Y_{1,1}(\bar z, z)=0\ .
\end{align}
\end{subequations}
Here we used orthornomality of the spherical harmonics on the disc, see Appendix~\ref{app:spherical harmonics}. We should also clarify that by $\mathrm{d}^2z$ we mean $\mathrm{d} \Re(z) \, \mathrm{d}\Im(z)$. We wrote the gauge condition as one complex condition here , which upon complex conjugation would also imply the vanishing of $X_{2,-1}$ and $X_{1,-1}$ respectively.

In order to show the admissibility, we define the complex-valued function
\be 
V(\alpha)=\int_D \frac{\mathrm{d}^2 z}{(1+|z|^2)^2} X \circ \gamma_\alpha^{-1} (z,\bar{z}) \overline{Y_{\ell,1}(z,\bar{z})}\ .
\ee
Note that $\overline{Y_{\ell,1}(z,\bar{z})}=Y_{\ell,-1}(\bar{z},z)$.
We will call it $V_{\mathrm{N}}(\alpha)$ when we set $\ell=1$ and we are dealing with Neumann boundary condition. Similarly for the Dirchlet case, we will call it $V_{\mathrm{D}}(\alpha)$  and set $\ell=2$. Showing admissibility of the gauge amounts to showing that $V(\alpha)$ has a zero in the unit disk. In fact, we should also determine the number of zeros since this will be needed in the calculation of the Faddeev-Popov determinant eventually.

It turns out that the number of zeros of $V(\alpha)$ in the unit disk can be determined from its behavior near the boundary by using Stokes' theorem as explained below. Thus, we first analyze the behavior of $V(\alpha)$ for $\alpha=\rho \, \mathrm{e}^{i \theta}$ and $\rho$ close to 1. This behavior of $V(\alpha)$ is entirely universal, because $\gamma_\alpha^{-1}(z)$ is close to the boundary of the worldsheet disk for any choice of $z$ and $\rho \sim 1$. Thus in this limit one is only probing the function $X$ close to the boundary of the worldsheet disk, where its behavior is specified by the boundary conditions.  We find
\begin{subequations} \label{eq:boundarybehave}
\begin{align}
V_{\mathrm{N}}(\alpha) &=i(1-\rho)e^{-i\theta} \underbrace{\sum_{\begin{subarray}{c} \ell \, m \\
\ell+m=\text{even}\end{subarray}} h_\text{N}(\ell,m)  \Im\left(X_{\ell,m} \mathrm{e}^{i m \theta}\right)}_{f_\text{N}(\theta)\equiv\, \text{real function}}\, +\, o(1-\rho)\ ,\\
V_{\mathrm{D}}(\alpha)&=(1-\rho)e^{-i\theta} \underbrace{\sum_{\begin{subarray}{c} \ell,\, m \\
\ell+m=\text{odd}\end{subarray}} h_\text{D}(\ell,m)  \Re\left(X_{\ell,m} \mathrm{e}^{i m \theta}\right)}_{f_\text{D}(\theta)\equiv\, \text{real function}}\,+\,o(1-\rho)\ .
\end{align}
\end{subequations}
The numbers $h_\text{N}(\ell,m)$ and $h_\text{D}(\ell,m)$ are real.
Eq.~\eqref{eq:boundarybehave} follows from the observation
\begin{equation}
\int_D \frac{4\, \mathrm{d}^2 z}{(1+|z|^2)^2} Y_{\ell,m} \circ \gamma_\alpha^{-1} (z,\bar{z}) \overline{Y_{1,1}(z,\bar{z})}=(1-\rho) e^{i(m-1)\theta} h_\text{N}(\ell,m) \,+\,o(1-\rho)
\end{equation}
with $h_\text{N}(\ell,m)=-h_\text{N}(\ell,-m)$ for the Neumann boundary condition. This leads to only the imaginary part of $X_{\ell,m} \mathrm{e}^{i m \theta}$ surviving in the sum. Furthermore, reality of $X_{0,0}$ implies the vanishing of the $m=0$ term. For Dirichlet boundary condition, we instead have
\begin{equation}
\int_D \frac{4\, \mathrm{d}^2 z}{(1+|z|^2)^2} Y_{\ell,m} \circ \gamma_\alpha^{-1} (z,\bar{z}) \overline{Y_{2,1}(z,\bar{z})}=(1-\rho) e^{i(m-1)\theta} h_\text{D}(\ell,m) \,+\,o(1-\rho)
\end{equation}
where $h_\text{D}(\ell,m)=h_\text{D}(\ell,-m)$. This leads to only the real part surviving. It is easy to compute these integrals in \texttt{Mathematica} for low values of $\ell$ and convince oneself of the validity of this behavior. We haven't tried to give a rigorous proof of this property.

Now we consider eq.~\eqref{eq:boundarybehave} and compute the following contour integral:
\be 
N\equiv \frac{1}{2\pi i}\int_{\partial D} \frac{\mathrm{d} V}{V}\ .
\ee
Here the contour encircles $D$ once in counterclockwise sense. To make this well-defined, we take the contour to be very close to the boundary. We can compute this directly from the behavior eq.~\!\eqref{eq:boundarybehave}:
\begin{equation}
N=\frac{1}{2\pi i} \int_0^{2\pi } \frac{\mathrm{d} (\mathrm{e}^{-i  \theta} f(\theta))}{\mathrm{e}^{-i  \theta}f(\theta)} =\frac{1}{2\pi i} \int_0^{2\pi } (-i \mathrm{d}\theta+\mathrm{d} \log f(\theta))=-1+w(f)\,.
\end{equation}
where $w(f)$ is the winding the number of the function $f(\theta)$ (which we called $f_\text{N}$ and $f_\text{D}$ in eq.~\eqref{eq:boundarybehave} depending on the boundary condition). We would like to conclude that the winding number $N$ of $V$ around the boundary is $-1$. However, $f(\theta)$ is real and is not generally sign definite, hence can potentially cross zero. For such functions, the winding number around zero is ill-defined. To cure this we perform the following replacement
\begin{subequations} \label{eq:shift}
\begin{align}
\text{Dirichlet}:&\qquad X\to X+ i\varepsilon Y_{1,0}\ ,\\
\text{Neumann}:&\qquad X\to X+ \varepsilon Y_{1,0}\ ,
\end{align}
\end{subequations}
with fixed $\varepsilon\ne 0$. This results in an additive modification of eq.~\!\eqref{eq:boundarybehave}; the modified function $f_{\mathrm{N}}(\theta)$ has a constant real piece while the modified $f_{\mathrm{D}}(\theta)$  has a constant imaginary piece. This guarantees that the modified function $f(\theta)$ does not pass through the origin and $w(f)=0$.  So with this modification, we have
\be\label{eq:gcs}
N =-1\ .
\ee
Before analyzing the above equation, let us discuss the meaning of the regularization.  The path integral can be understood as a contour integral in the space of complex-valued $L^2$-functions. This translates into the reality condition $\overline{X_{\ell,m}}=X_{\ell,-m}$ which specifies the contour for the modes. However, one can slightly shift the contour which should leave the value of the path integral unchanged. For the Dirichlet case,  the eq.~\!\eqref{eq:shift} amounts to $X_{1,0}\to X_{1,0}+i \varepsilon$. This should be thought of as doing the Gaussian integral over the $\mathrm{Im}X_{1,0}=\varepsilon$ line instead of on the real line.\footnote{The interpretation for the Neumann case is not as simple as the Dirichlet one, since here we are regulating using a component which does not really respect the Neumann boundary condition.}
We should also mention that the details of this modification do not matter. We could modify $X$ in any infinitesimal way, since any generic perturbation of a real function will result in a vanishing winding number. We just choose \eqref{eq:shift} for definiteness.

Eq.~\!\eqref{eq:gcs} implies that $V$ has exactly one zero in the disk, provided one counts zeros with signs and multiplicities as follows. For a generic complex function $V$ on the unit disk, zeros are isolated. We can encircle a zero by a contour and view $V(\alpha)$ restricted to the contour as a map $\mathrm{S}^1 \longmapsto \mathbb{C} \setminus \{0\}$. There is a winding number associated to this map which is the order of zero. For example the function $V(\alpha)=\alpha$ has a zero of order 1 around the origin, whereas the function $V(\alpha)=\bar{\alpha}$ has a zero of order $-1$ around the origin. For a zero of order $n$, we compute easily
\be 
\int_\mathcal{C} \frac{\mathrm{d}V}{V}=n\ ,
\ee
where the contour $\mathcal{C}$ encircles only the zero of $V$. Now by Stokes' theorem it follows that the sum of the orders of zeros has to be $-1$. In particular, there is at least one zero and the gauge is admissible. The significance of minus sign will become clear in following section when we discuss a signed version Faddeev-Popov gauge fixing procedure.

Once we have proved that the gauge is admissible, the regularization parameter $\varepsilon$ does not matter and can be set to $0$. We will do so in rest of the calculation.

For different gauges with where we impose $X_{\ell,m}=0$ with $m\not\in \{-1, 1\}$, we should instead consider 
\be 
V(\alpha)=\int_D \frac{\mathrm{d}^2 z}{(1+|z|^2)^2} X \circ \gamma_\alpha^{-1} (z,\bar{z}) \overline{Y_{\ell,m}(z,\bar{z})}\ .
\ee
and then the overall winding number $N$ turns out to be $-m$. In what follows we will use the gauge where $m=1$. It is possible to perform the computation with other choice of gauge as well with $m\neq 1$ (as long as $m \ne 0$ in which the gauge is no longer admissible).

\subsection{Computation of the path integral}
After these preparations, the actual computation of the gauge-fixed partition function is very easy. We can apply the modified Faddeev-Popov procedure that we reviewed in Appendix~\ref{app:FP procedure} to our problem. It is modified in that it counts intersections of the gauge orbit with the gauge slice with signs. This is necessary because while the gauge we have chosen is admissible, it is not uniquely so. The modified FP-procedure cancels unwanted intersections of the gauge orbit and the gauge slice by counting them with minus signs. The gauge group is $\mathrm{PSL}(2,\mathbb{R})$ and the gauge condition is $F(X)=(X^g)_{1,1}=0$ for Neumann and $F(X)=(X^g)_{2,1}=0$ for Dirichlet boundary conditions. The computation in the previous Section~\ref{subsec:gauge choice} shows in fact precisely that the intersection number $\mathcal{I}$ between the gauge orbit and the gauge slice is $\mathcal{I}=-1$, independent of $X$, i.e.\ 
\begin{equation}
-1=\int_\mathcal{G} \mathrm{d}g\  \mathop{\text{det}} \mathop{\text{Jac}} F(X^g)\, \delta(F(X^g))\ .
\end{equation}
For $m\neq 1$, the LHS of the above equation reads $-m$ instead of $-1$, since the intersection number is $\mathcal{I}=-m$. In what follows we will use $m=1$.

\paragraph{Neumann Condition.}
The Neumann condition involves the modes with $\ell+m$ even. The gauge fixing condition is $F(X)=X_{1,1}=0$. The Jacobian 
\be 
\mathop{\text{Jac}} F(X^g)
\ee
is linear in $X$. Hence it can be evaluated mode by mode. It is actually only non-vanishing for finitely many values of mode $X_{\ell,m}$. 
When expressing the group element $g$ in terms of $\alpha \in \mathrm{PSU}(1,1)/\mathrm{U}(1)$ through \eqref{eq:gammaalpha} (and writing $X^{\gamma_\alpha} \equiv X^\alpha$), we have in fact the identity
\begin{equation}
1=-\int\frac{\pi\, \mathrm{d}^2\alpha}{(1-|\alpha|^2)^2}\  J_{\mathrm{N}}(X^\alpha)\ \delta^2( F(X^{\alpha}))
\end{equation}
with
\begin{equation}
\pi J_{\mathrm{N}}(X)=\frac{36 }{5}(\mathrm{Im}X_{2,2})^2+\frac{36 }{5}(\mathrm{Re}X_{2,2})^2-\frac{6}{5} X_{2,0}^2\ .
\end{equation}
The gauge-fixed path integral hence reads explicitly
\be
Z^{\mathrm{N}}_{\text{disk}}=-\int\mathscr{D}X\ \delta(\mathrm{Re}X_{1,1}) \delta(\mathrm{Im}X_{1,1})\, J_{\mathrm{N}}(X) \, \mathrm{e}^{-S[X]}\ .
\ee
where the action in terms of modes is given by
\be 
S[X]=\frac{1}{4\pi\alpha'} \sum_{\ell+m\in 2\mathbb{Z}}\ell(\ell+1)|X_{\ell,m}|^2\ .
\ee
Hence in the ratio of the gauged and the ungauged CFT partition all but finitely many modes cancel. Thus it is given by a simple ratio of Gaussian integrals. It works out to be
\begin{align}
\frac{Z^{\mathrm{N}}_{\text{disk}}}{Z^{\mathrm{N}}_{\text{CFT}}}&=-\frac{2}{\pi^2}\ .
\end{align}
\paragraph{Dirichlet boundary conditions.} The computation is completely analogous. The Fadeev-Popov determinant works out to be
\be 
\pi J_{\mathrm{D}}(X)=\frac{64 }{7}\left[(\Im X_{3,2})^2+(\Re X_{3,2})^2\right]-\frac{16}{5} \sqrt{\frac{3}{7}} X_{1,0} X_{3,0}-\frac{2 }{5}(X_{1,0})^2-\frac{96}{35}(X_{3,0})^2 
\ee
in this case. In particular it again only involves finitely many modes and allows one to reduce the ratio of the gauged and the ungauged partition function to a ratio of finite-dimensional integrals. One again recovers
\begin{tcolorbox}
\be
\frac{Z_{\text{disk}}^\text{D}}{Z_{\text{CFT}}^\text{D}}=\frac{Z_{\text{disk}}^\text{N}}{Z_{\text{CFT}}^\text{N}}= -\frac{2}{\pi^2}=\text{Regularized volume of}\ \mathrm{PSL}(2,\mathbb{R})\  ,
\ee
\end{tcolorbox}
in agreement with the regularization procedure discussed in  \cite{Liu:1987nz}. This is the result we anticipated in eq.~\!\eqref{eq:Main}.

\section{Gauge fixing \texorpdfstring{$\boldsymbol{\mathrm{d}X(0)=0}$}{dX(0)=0}} \label{sec:alternative gauge}
In this section, we repeat the calculation using a different gauge choice. We mostly focus on the Neumann case and indicate the necessary changes for the Dirichlet case. We used the gauge choice $X_{1,\pm 1}=0$ before. The difficulty for this gauge choice was to establish admissibility. We saw that the gauge is not uniquely fixed, but counting solutions with a sign of the corresponding Jacobian that enters the Faddeev-Popov determinant, there is always a unique solution (up to the subtlety that we had to shift the contour slightly in the complex plane). On the other hand, it was almost trivial to compute the path integral with the insertion of the corresponding delta-function and the Jacobian, because this only involved finitely many modes $X_{\ell,m}$.

In this section we will shift the difficulty -- our gauge choice is easily seen to be admissible, but computing the actual path integral will be more technical.
\subsection{Admissibility and uniqueness}
Our gauge condition reads
\be 
\mathrm{d} X(0)=0\ ,
\ee
i.e.~the center of the disk is a local extremum for one of the spacetime coordinates $X$. As before, this leaves the $\mathrm{U}(1) \subset \mathrm{PSL}(2,\mathbb{R})$ subgroup unbroken. But since $\mathrm{U}(1)$ is compact, it simply yields an additional factor of $\pi$ in the final result.\footnote{The volume of $\mathrm{U}(1)$ is $\pi$ and not $2\pi$ because the gauge group is $\mathrm{PSL}(2,\mathbb{R})$ and not $\mathrm{SL}(2,\mathbb{R})$.} We will first discuss this condition for Neumann boundary conditions.

Before discussing admissibility of this gauge, we should address a subtlety. The restriction $X|_{\partial D}$ is a function on $\partial D \cong \mathrm{S}^1$ and as such would have local extrema (at least two of them). Since for Neumann boundary conditions, also $\partial_n X|_{\partial D}=0$, it follows that this local extrema of $X|_{\partial D}$ are also local extrema of $X$. Thus for generic $X$ there are always local extrema on the boundary of the disk. This is undesirable for our purposes.
To rectify this behavior, we modify slightly the boundary condition as follows:
\be 
\partial_n X(z)\Big|_{\partial D}=\varepsilon
\ee
for small $\varepsilon$. $\varepsilon$ can in principle be a non-trivial function on the boundary of the disk -- our only requirement is that it doesn't possess a zero. We think of $\varepsilon$ as being very small. This choice guarantees us that there will be no local extrema on the boundary of the disk. Our modification either shifted them slightly outside or inside of the disk. 

\begin{figure}[!ht]
\begin{center}
\includegraphics[width=.5\textwidth]{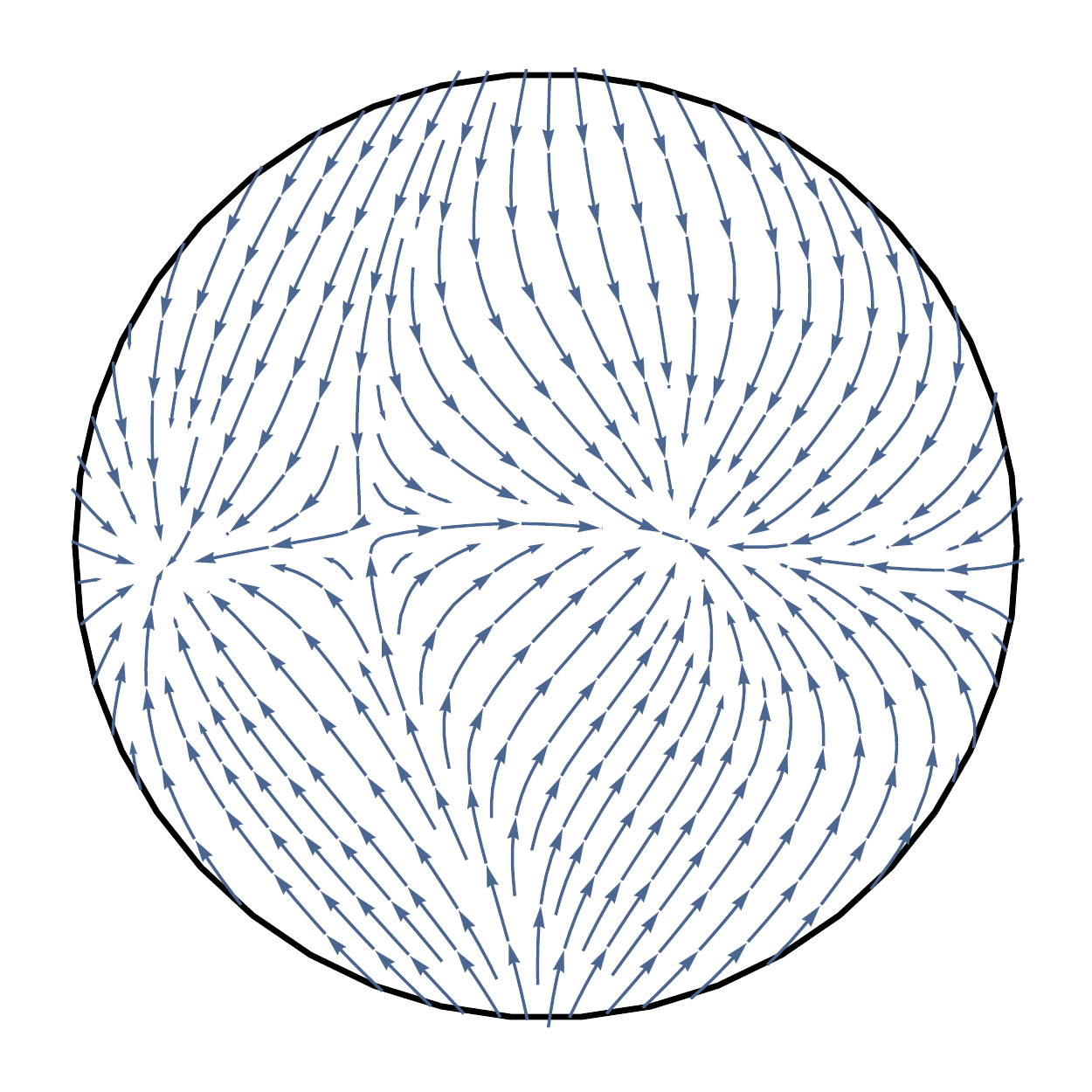}
\end{center}
\caption{The derivative $\mathrm{d}X$ on the disk.} \label{fig:vectorfieldN}
\end{figure}

Now we can discuss admissibility of the gauge. For this consider $\mathrm{d}X$, which we can view as a vectorfield over $D$. We equip $D$ with a flat metric, so that vectorfields can be identified with 1-forms. Then this vectorfield has roughly the form as depicted in figure~\ref{fig:vectorfieldN}. In the example of the figure, there are three extrema: two (local) maxima and one saddle point. 
Thus, our gauge choice is admissible in this example, but not uniquely so. In general, the number of (local) maxima, minima and saddlepoints is constrained by the Poincar\'e-Hopf theorem.\footnote{Or alternatively by the Morse lemma when $X$ is a Morse function.} The Poincar\'e-Hopf theorem says that for a vectorfield of the form we are considering
\be \label{eq:topocons}
\text{\# maxima}-\text{\# saddle points}+\text{\# minima}=1\ .
\ee
The RHS of this equation is the Euler characteristic of the disk. This equation shows in particular that the gauge is admissible.

We are thus in a similar situation as for the other gauge, where the gauge is not uniquely fixed, but different solutions to the gauge condition are constrained by a topological condition. We can exploit this by considering the following quantity
\be 
\int_{\mathrm{PSL}(2,\mathbb{R})} \mathrm{d}\gamma\  \det(\text{Hess}(X^\gamma)(0)) \delta^2(\mathrm{d} X^\gamma(0))\,.
\ee
 Here, $\mathrm{d}\gamma$ is the Haar measure and $X^\gamma \equiv X \circ \gamma^{-1}$ as before.  $\text{Hess}(X)(0)$ is the Hessian matrix
\be 
\text{Hess}(X)(0)=\begin{pmatrix}
\partial_x^2 X(0) & \partial_x \partial_y X(0)\\
\partial_x \partial_y X(0) & \partial_y^2 X(0)
\end{pmatrix}\ .
\ee
Given our previous discussion, we can evaluate this expression very explicitly. As before, we can parametrize the coset $\mathrm{PSL}(2,\mathbb{R})/\mathrm{U}(1)$ by $\alpha \in D$, see eq.~\eqref{eq:gammaalpha}. Following the logic of the modified Faddeev-Popov procedure, this evaluates to
\begin{align}
\int_D \frac{\pi \, \mathrm{d}\alpha\, \mathrm{d}\bar{\alpha}}{(1-|\alpha|^2)^2} \det(\mathrm{Hess}(X^\alpha)(0)) \delta^2(\mathrm{d}X^\alpha(0))=\pi \sum_{\alpha_0}  \text{sgn}\left(\det \text{Hess}(X(\alpha_0))\right)\ .
\end{align}
We finally have
\begin{align} 
\text{sgn}\left(\det \text{Hess}(X(\alpha_0))\right)=\begin{cases}
+1 & \text{$\alpha_0$ is a maximum or minimum of $X(z)$} \\
-1 & \text{$\alpha_0$ is a saddlepoint of $X(z)$}
\end{cases}
\end{align}
Thus, by the topological constraint \eqref{eq:topocons} on the maxima, minima and saddlepoints, we have simply
\begin{align}
\int_D \frac{\pi \, \mathrm{d}\alpha\, \mathrm{d}\bar{\alpha}}{(1-|\alpha|^2)^2} \det(\mathrm{Hess}(X^\alpha)(0)) \delta^2(\mathrm{d}X^\alpha(0))=\pi \ .
\end{align}
In other words, the intersection number between the gauge slice and the gauge orbit is $\mathcal{I}=1$. The general logic is again given by the modified FP-procedure that we review in Appendix~\ref{app:FP procedure}. We finally insert this identity in the path integral for the disk partition function
\begin{multline}
\int \frac{\mathscr{D}X}{\mathop{\text{vol}}\mathrm{PSL}(2,\mathbb{R}) }\mathrm{e}^{-S[X]}\\
=\frac{1}{\pi} \int \frac{\mathscr{D}X}{\mathop{\text{vol}}\mathrm{PSL}(2,\mathbb{R}) } \int_D \frac{\pi \, \mathrm{d}\alpha\, \mathrm{d}\bar{\alpha}}{(1-|\alpha|^2)^2} \det(\mathrm{Hess}(X^\alpha)(0)) \delta^2(\mathrm{d}X^\alpha(0))\mathrm{e}^{-S[X]}\ .
\end{multline}
While we suppressed the other directions of the sigma model as well as the ghost fields, we should remember that they are present in order to have a non-anomalous $\mathrm{PSL}(2,\mathbb{R})$ symmetry. We suppress them from the notation for simplicity. With this convention, both the measure and the action are invariant under $\mathrm{PSL}(2,\mathbb{R})$ transformations -- $\mathscr{D}X=\mathscr{D}X^\gamma$ and $S[X^\gamma]=S[X]$. Thus, after replacing $X$ by $X^\alpha$ in the measure and the action, we can rename $X^\alpha \to X$ everywhere. The $\alpha$-integral then formally is
\be 
\int_D \frac{\pi \, \mathrm{d}\alpha\, \mathrm{d}\bar{\alpha}}{(1-|\alpha|^2)^2}=\int_{\mathrm{PSL}(2,\mathbb{R})} \mathrm{d}\gamma=\mathop{\text{vol}}\mathrm{PSL}(2,\mathbb{R}) \ ,
\ee
which cancels the corresponding factor in the denominator (at least this is our definition what we mean by $\mathop{\text{vol}}\mathrm{PSL}(2,\mathbb{R}) $). Thus, we end up with the following gauge-fixed form of the disk partition function
\be 
Z_\text{disk}=\frac{1}{\pi} \int \mathscr{D}X \det(\mathrm{Hess}(X)(0)) \delta^2(\mathrm{d}X(0))\mathrm{e}^{-S[X]}\ . \label{eq:disk partition function gauge fixed 2}
\ee
\paragraph{Dirichlet case.} Let us indicate the changes for the Dirichlet case. Here, $X|_{\partial D}=0$ and so the derivative along the boundary of $X$ vanishes. Hence we again expect that generically there can be critical points of $X(z)$ on the boundary $\partial D$ and we require a similar regularization as before. This situation is topologically completely equivalent to the Neumann case if we rotate the vectorfield pointwise by 90 degrees. Then the normal derivative and the derivative along the boundary get interchanged and we are back to the Neumann situation that can be regularized as discussed above. Thus, we again have after regularization
\be 
\text{\# maxima}-\text{\# saddle points}+\text{\# minima}=1\ .
\ee
The rest of the computation did not require the boundary condition and hence \eqref{eq:disk partition function gauge fixed 2} also holds for Dirichlet boundary conditions.
\subsection{Computation of the path integral}
Next, we compute the gauge-fixed path integral eq.~\!\eqref{eq:disk partition function gauge fixed 2}. We choose a flat metric on the disk for simplicity and set $\alpha'=1$. We will again perform the computation first for Neumann boundary conditions and indicate the changes for Dirichlet boundary conditions below. Let us introduce the standard generating functional
\be 
W(J)=\left \langle \exp\left(i \int \mathrm{d}^2z\ X(z) J(z) \right) \right \rangle\ ,
\ee
where the correlation function is normalized such that $\langle 1 \rangle=1$. Here, $J(z)$ is an arbitrary source for $X$. We can compute the generating functional in the following standard way. The Green's function for the Laplacian on the disk with Neumann boundary conditions reads
\be 
G(z,w)=\frac{1}{2\pi}\left(\log |z-w|+\log \left(|w||z-w^*|\right)\right)-\frac{1}{4\pi}(|z|^2+|w|^2)\ ,
\ee
where $w^*=\frac{w}{|w|^2}$ is the point reflected at the unit circle. This Green's function is symmetric, which becomes obvious if we write it in the form
\be 
G(z,w)=\frac{1}{2\pi}\left(\log |z-w|+\log |1-z \bar{w}|\right)-\frac{1}{4\pi}(|z|^2+|w|^2)\ . \label{eq:Greens function N}
\ee
It satisfies
\be 
\Delta_z G(z,w)=\delta^2(z,w)-\frac{1}{\pi}\ .
\ee
The correction is expected, because the Laplacian has a zero mode and thus the inverse only exists for non-zero modes. 
One can complete the square in the path integral and derive
\be 
W(J)=\exp\left(\pi \int \mathrm{d}^2 z \ \mathrm{d}^2 w\ G(z,w) J(z) J(w) \right)\ .
\ee
This expression is valid as long as the zero mode $\int \mathrm{d}^2 z\ J(z)$ vanishes. This will always be satisfied below since our gauge fixing condition does not involve the zero mode.

Now we turn again to eq.~\!\eqref{eq:disk partition function gauge fixed 2}. 
It involves composite operators such as the determinant of the Hessian which have to be defined properly. Our regularization is to use point splitting.
Correspondingly, the determinant of the Hessian becomes
\be 
\partial_x^2 X(z_x)\partial_y^2 X(z_y)-\partial_x \partial_y X(z_x)\partial_x \partial_y X(z_y)\ .
\ee
Here and in the following $\partial_x$ ($\partial_y$) is the derivative with respect to the real (imaginary) part of the complex argument. We find it less confusing to use real coordinates in the computation. We used $z_x$ and $z_y$ for the two point-split points to remember which one carries more $x$ and $y$-derivatives. We ultimately want to take them both to zero. Similarly, the $\delta$-functions can be taken to be
\be 
\delta(\partial_x X(z_x))\delta(\partial_y X(x_y)) \ .
\ee
It turns out that in the following computation it is very natural to take them at the same coordinates as the entries of the Hessian matrix -- this will not lead to singularities. In fact, this point-split version of the integral simply comes from the modified gauge condition
\be 
\partial_x X(z_x)=0\quad \text{and}\quad \partial_y X(z_y)=0\ .
\ee
As a first step, we can compute
\begin{align} 
\tilde{W}(J)&=\left\langle \delta(\partial_x X(z_x))\delta(\partial_y X(z_y)) \exp\left(i \int \mathrm{d}^2z\ X(z) J(z) \right) \right \rangle \\
&=\frac{1}{(2\pi)^2} \int _{-\infty}^\infty \mathrm{d}k_x \mathrm{d} k_y \ W\big(J+k_x \partial_x \delta^2(z-z_x)+k_y \partial_y \delta^2(z_y)\big)\ .
\end{align}
Notice that as promised, the modified source still does not have a zero mode. We can plug in the explicit form of $W(J)$ to obtain
\begin{align}
\tilde{W}(J)&=\frac{W(J) }{(2\pi)^2} \int _{-\infty}^\infty \mathrm{d}k_x \mathrm{d} k_y \ \exp\Bigg(\pi \sum_{i,j \in \{x,y\}} k_i k_j\partial_i^{(1)}\partial_j^{(2)}G(z_i,z_j)\nonumber\\
&\qquad\qquad\qquad-2\pi\sum_{i\in \{x,y\}}k_i\int \mathrm{d}^2 z \ \partial_{i}^{(2)}G(z,z_i) J(z)\Bigg)\ .
\end{align}
The superscript $(1)$ and $(2)$ indicates whether the derivative acts on the first or second entry of the Green's function.
Remembering that we use point splitting to define Green's functions at coincident points, we need to subtract the singular piece of the Green's function that is  $\frac{1}{2\pi} \log |z-z|$. This gives
\be 
G_\text{reg}(z,z)=\frac{1}{2\pi}\left(\log\left(1-|z|^2\right)-|z|^2\right)\ .
\ee
We next compute the integral over $k_x$ and $k_y$. Let
\be 
A_{i,j}=-\partial_i^{(1)}\partial_j^{(2)}G(z_i,z_j)\ , \qquad b_i= \int \mathrm{d}^2 z \ \partial_{i}^{(2)}G(z,z_i) J(z)\ .
\ee
We thus simply compute the Gaussian integral with the result
\be 
\tilde{W}(J)=\frac{W(J) }{(2\pi)^2\sqrt{\det(A)}}\exp\left(\pi \sum_{i,j} b_i (A^{-1})_{i,j} b_j\right)\ .
\ee
It turns out that the matrix $A$, although complicated is indeed positive definite so that the integral over $k_x$ and $k_y$ is well-defined. By direct computation, we have
\be 
\det(A)\Big|_{z_x=0,z_y=0}=\frac{1}{(2\pi)^2}\ .
\ee
Also the exponential behaves nicely in the limit where $z_x \to 0$ and $z_y \to 0$ and we obtain
\be 
\sum_{i,j} b_i (A^{-1})_{i,j} b_j=2\pi \int\mathrm{d}^2 z\ \mathrm{d}^2 w\ \sum_{p\in \{x,y\}}\partial_p^{(2)} G(z,0)\partial_p^{(2)} G(w,0) J(z) J(w) 
\ee
Let us define
\be 
\tilde{G}(z,w)=G(z,w)+2\pi\sum_{i\in \{x,y\}}\partial_i^{(2)} G(z,0)\partial_i^{(2)} G(w,0)\ . \label{eq:G tilde}
\ee
Thus, after specialization of $z_x=z_y=0$, we have
\be 
\tilde{W}(J)=\frac{1}{2\pi} \exp\left(\pi \int \mathrm{d}^2 z \ \mathrm{d}^2 w\ \tilde{G}(z,w) J(z) J(w) \right)
\ee
To complete the computation, we also want to include the effect of the Hessian. Point-splitting again, we simply obtain it by taking functional derivatives. Remembering also the additional factor of $\frac{1}{\pi}$ from the volume of the residual gauge group $\mathrm{U}(1)$, we want to compute
\begin{align} 
\frac{Z_\text{disk}}{Z_\text{CFT}}&=-\frac{1}{2\pi^2}\lim_{z_x \to 0,\, z_y \to 0} \left((\partial_x^{(1)})^2 (\partial_y^{(2)} )^2-\partial_x^{(1)}\partial_y^{(1)}\partial_x^{(2)}\partial_y^{(2)}\right)\frac{\delta}{\delta J(z_x)}  \frac{\delta}{\delta J(z_y)}  \tilde{W}(J) \Big|_{J=0}\\
&=-\frac{1}{\pi}\lim_{z_x \to 0,\, z_y \to 0} \left((\partial_x^{(1)})^2 (\partial_y^{(2)} )^2-\partial_x^{(1)}\partial_y^{(1)}\partial_x^{(2)}\partial_y^{(2)}\right)\tilde{G}(z_x,z_y)\ .
\end{align}
Here, $Z_\text{CFT}$ is the CFT partition function without gauging of $\mathrm{PSL}(2,\mathbb{R})$.
There are two terms -- from the original $G(z,w)$ and from the correction term in eq.~\!\eqref{eq:G tilde}. The second term leads again to Green's functions at coincident points which we regularize as before. 
A direct computation then leads to
\be 
\frac{Z_\text{disk}}{Z_\text{CFT}}=-\frac{1}{\pi} \times \frac{2}{\pi}=-\frac{2}{\pi^2}\ .
\ee
This is in perfect agreement with out earlier calculation.
\paragraph{Dirichlet case.} For Dirichlet boundary conditions, the following changes need to be made. The Green's function now takes the form
\be 
G(z,w)=\frac{1}{2\pi}\left(\log |z-w|-\log |1-z \bar{w}|\right)  \label{eq:Greens function D}
\ee
and there is no zero mode. Furthermore, the matrix $A_{i,j}$ is \emph{negative definite} in this case and thus the integral over $k_x$ and $k_y$ is a priori ill-defined. However, one can still go on by employing a double Wick rotation $k_p \to i k_p$ (but the answer is less well-defined in this case). This leads to
\be 
\tilde{W}(J)=-\frac{W(J) }{(2\pi)^2\sqrt{\det(A)}}\exp\left(\pi \sum_{i,j} b_i (A^{-1})_{i,j} b_j\right)\ ,
\ee
where the various quantities are given by analogous expressions as in the Neumann case. The extra minus sign comes from the analytic continuation. The Wick rotation exchanges branches of the square root. The remaining steps are completely analogous and one obtains the result
\be 
\frac{Z_\text{disk}}{Z_\text{CFT}}=\frac{1}{\pi} \times \left(-\frac{2}{\pi}\right)=-\frac{2}{\pi^2}\ .
\ee

\section{Relation to a one-point function} \label{sec:one point function}
In this section, we will explain yet another method to compute the disk partition function by relating it to a one-point function. This is more along the lines how the disk partition functions were evaluated previously in the literature. Actually, this was done historically by using the soft dilaton theorem \cite{Shapiro:1975cz, Ademollo:1975pf} that relates the disk partition function to a one-point function of the dilaton with zero momentum. This exploits the fact that the dilaton appears in the spacetime effective action as an exponential. The computation we present here is simpler because one does not have to deal with the subtleties of the dilaton vertex operator and one does not have to make any assumption about the spacetime theory.

\subsection{Marginal operator}
Let us suppose that there is a circle of radius $R$ in the spacetime which is described by a compact free boson $X \sim X+2\pi L$.
As before, we want to compute the path integral over the worldsheet CFT with a $\mathrm{PSL}(2,\mathbb{R})$ gauging  and compare it with the path integral without gauging. 

We make use of the fact that the worldsheet partition function as well as the gauged string partition function should behave in a simple way on $L$. In fact, $L$ only enters in the path integral formalism through the zero modes which leads to the behavior
\begin{subequations}
\begin{align} 
\text{Neumann}&:\  Z_\text{CFT} \propto L^1\ , \\
\text{Dirichlet}&:\  Z_\text{CFT} \propto L^0\ ,
\end{align} \label{eq:proportionalities}
\end{subequations}
because the zero mode is only present for the Neumann boundary condition. We assume that this property continues to be true in the full string partition function $Z_\text{disk}$.

In the worldsheet path integral
\be 
Z_\text{CFT}=\int \mathscr{D}X \ \mathrm{e}^{-S[X]}\ ,
\ee
we can make the $L$-dependence explicit by defining $X'=L^{-1} X$, which has the periodicity. Then the worldsheet path integral reads
\be 
Z_\text{CFT}=L^\gamma\int \mathscr{D}X' \ \mathrm{e}^{-L^2 S[X']}\ ,
\ee
We put a prefactor $L^\gamma$ in front of the path integral to account for the fact that the measure $\mathscr{D}X'$ should also transform under this replacement. Since the replacement $X'=L^{-1} X$ is linear, the most general transformation is given by an overall factor $L^\gamma$. However, the precise value of the exponent $\gamma$ is scheme dependent and we leave it open. One can for example compute that in zeta-function regularization $\gamma=\frac{1}{6}$. Let us write $V(z) =g^{ab} \partial_a X' \partial_b X'(z)$ in the following for simplicity. We thus have
\be 
\frac{\partial_L (L^{-\gamma} Z_\text{CFT})}{L^{-\gamma} Z_\text{CFT}}=-\frac{L}{2\pi \alpha'} \frac{\int \mathscr{D}X' \ \int \mathrm{d}^2z \ \sqrt{g} \, V(z) \mathrm{e}^{-L^2 S[X']}}{\int \mathscr{D}X' \ \mathrm{e}^{-L^2 S[X']}}
\ee
In this expression it is now very simple to gauge fix because we are computing a one-point function. We can put the vertex operator $V(z)$ in the center of the disk. We take the disk again to be the unit disk with flat metric so that the vertex operator is inserted at $z=0$. The remaining Faddeev-Popov determinant is simply $\frac{1}{\pi}$ coming from the unbroken $\mathrm{U}(1)$. We thus deduce
\be 
\frac{\partial_L (L^{-\gamma} Z_\text{disk})}{L^{-\gamma} Z_\text{CFT}}= -\frac{L}{2\pi^2 \alpha'} \langle V(0) \rangle_L\ ,
\ee
where the normalized expectation value is taken w.r.t.\ the action $L^2 S[X']$.
\subsection{Computation}
After having related the disk partition function to a one-point function, we proceed with the calculation. The expectation value $\langle V(0) \rangle_L$ can be computed via Green's functions as in Section~\ref{sec:alternative gauge}. To start, we first point split the operator $V(z)$ and compute the two point function
\be 
4 \langle \partial X(z) \bar{\partial}X(w) \rangle
\ee
instead which in the limit $z,w \to 0$ gives the desired one-point function. Here we wrote again $X$ for $X'$ to avoid cluttering the notation.
This gives
\be 
\frac{\partial_L (L^{-\gamma} Z_\text{disk})}{L^{-\gamma} Z_\text{CFT}}=-\frac{L}{2\pi^2 \alpha'} \times \left(-\frac{2\pi \alpha'}{L^2} \right) \times 4 \lim_{z,w \to 0}\partial_z \bar{\partial}_w G(z,w)\ .
\ee
The additional factor comes from the generating functional $W(J)$ that we determine as in Section~\ref{sec:alternative gauge}. 

Notice that so far everything works with both boundary conditions. We also make the important remark that through point-splitting we have chosen a renormalization scheme and thus we can only expect agreement for a specific $\gamma$. For this reason we will consider a combination of the Neumann and Dirichlet partition functions where the scheme dependence cancels. We can compute the ratio
\be 
\frac{Z_\text{disk}}{L Z_\text{CFT}}= \frac{\partial_L (L^{-\gamma} Z_\text{disk}^\text{N})}{L^{-\gamma} Z_\text{CFT}^\mathrm{N}}-\frac{\partial_L (L^{-\gamma} Z_\text{disk}^\text{D})}{L^{-\gamma} Z_\text{CFT}^\mathrm{D}}\ .
\ee
In this equality, we used the proportionalities \eqref{eq:proportionalities} as well as the expectation that the ratio $Z_\text{disk}/Z_\text{CFT}$ does not depend on the boundary conditions as well as independent of $L$. We finally learn
\begin{align} 
\frac{Z_\text{disk}}{Z_\text{CFT}}&=\frac{4}{\pi} \lim_{z,w \to 0} \partial_z \bar{\partial}_w \left(G^\text{N}(z,w)-G^\text{D}(z,w)\right) \\
&=\frac{4}{\pi} \lim_{z,w \to 0} \partial_z \bar{\partial}_w \left(\frac{1}{\pi} \log |1-z \bar{w}|-\frac{1}{4\pi} (|z|^2+|w|^2)\right) \\
&=-\frac{2}{\pi^2} \lim_{z,w \to 0} \frac{1}{(1-z \bar{w})^2}=-\frac{2}{\pi^2}\ ,
\end{align}
in agreement with our previous results. Here we used the explicit form of the Green's function eq.~\!\eqref{eq:Greens function N} and eq.~\!\eqref{eq:Greens function D}.

\section{Application to D-branes} \label{sec:D brane tension}
In this section, we apply our method to the computation of D-brane tension. Let us imagine a setup with a D$p$-brane in directions $0$ through $p$ (in flat spacetime). Then without turning on any fluxes, the worldvolume action of the D-brane is given by the DBI-action -- the higher-dimensional generalization of the Nambu-Goto action (in the Einstein frame):
\be 
S_{\text{D$p$-brane}}=T_p \int \mathrm{d}^{p+1} x\ \sqrt{\det (G^{(p)})}=T_p \vol(\mathrm{D}p)\ ,
\ee
where $\vol(\mathrm{D}p)$ is the $(p+1)$-dimensional worldvolume in spacetime the D-brane occupies and $T_p$ is the D$p$-brane tension -- the object we want to compute. We do not turn on any $B$-field or gauge field background values. The fact that $\vol(\mathrm{D}p)$ is infinite is not a problem in our analysis. We could imagine that in a Euclidean spacetime, directions $0$ through $p$ are toroidally compactified so that the worldvolume becomes finite. We already know that $T_p \propto g_\text{s}^{-1}$ (the closed string coupling) since D-branes are non-perturbative objects. Hence the partition function of the system is to leading order in $g_s$ given by
\be 
Z_{\text{D$p$-brane}}=\mathrm{e}^{-S_{\text{D$p$-brane}}}=\mathrm{e}^{-T_p \vol(\mathrm{D}p)}\ 
\ee
This partition function needs to be reproduced by a worldsheet computation. To leading order in $g_\text{s}$, the worldsheet partition function of a single open string ending on the D-brane is given by the disk partition function $Z_\text{disk}$. To account for the fact that there can be arbitrarily many strings present we need to exponentiate the single-string answer. So we require
\be 
\mathrm{e}^{-T_p \vol(\mathrm{D}p)} \overset{!}{=}\mathrm{e}^{Z_\text{disk}+\mathcal{O}(1)}\ .
\ee
Hence
\be 
T_p=-\frac{Z_\text{disk}}{\vol(\mathrm{D}p)}= -\frac{Z_\text{CFT}^{(p)}}{\vol(\mathrm{D}p)\vol(\mathrm{PSL}(2,\mathbb{R}))}\  .
\ee
Here we used the above computations that showed that passing from the disk partition function with $\mathrm{PSL}(2,\mathbb{R})$ gauged to the ungauged CFT partition function gives rise to a relative factor given by the effective volume of $\mathrm{PSL}(2,\mathbb{R})$. The superscript $(p)$ reminds us that there are $p+1$ Neumann directions and $D-p-1=25-p$ Dirichlet directions in the partition function.

We also note that it was crucial that the effective volume of $\mathrm{PSL}(2,\mathbb{R})$ turned out to be negative in order to get a positive D-brane tension.\footnote{One could repeat the same computation for O-planes, whose tensions are computed by the projective plane $\mathbb{RP}^2$ diagram. In this case, the residual symmetry group is $\mathrm{SO}(3)$, which is compact. Correspondingly, the tension of O-planes turns out to be \emph{negative}.}
\subsection{\texorpdfstring{$p$-dependence}{p-dependence}} 
As a first step in out computation, we fix the $p$-dependence of $T_p$. We use the fact that the effective volume of $\mathrm{PSL}(2,\mathbb{R})$ can be assigned a finite regularized value (the precise value becomes important only in the next subsection) and arrive at
\be 
\frac{T_{p+1}}{T_p}=\frac{Z_\text{CFT}^{(p+1)}}{Z_\text{CFT}^{(p)}\vol(\mathbb{R})}=\frac{Z_\text{CFT}^\text{N}}{Z_\text{CFT}^\text{D}\vol(\mathbb{R})}\ ,
\ee
where $Z_\text{CFT}^\text{N,D}$ are the CFT partition functions for a single free boson. All other directions in the worldsheet partition function as well as the ghost partition functions cancel. The volume appearing here is the volume in the direction $p+1$. This will remove the zero mode from the Neumann partition function. Let us compute the partition function on a hemisphere of radius $R$ in zeta-function renormalization \cite{Hawking:1976ja}.  
The non-zero modes lead to
\be 
Z_\text{CFT}^\text{N,D}=\text{(zero modes)}\times\prod_{\lambda} \sqrt{\frac{4\pi^2 \alpha' R^2}{\lambda}}\ .
\ee
The product runs over all eigenvalues of $-\Delta$ on the unit sphere with the correct boundary conditions. 
The zero mode for the Neumann condition leads to the following contribution. By definition, we normalized the path integral as follows. Choose an orthonormal basis of $\Delta$. Then the path integral is simply given by the usual integral over the all the coefficients in this orthonormal basis. The constant function is hence normalized as $\frac{1}{\sqrt{2\pi} R}$. Thus, the zero mode integral is
\be 
\int_{-L\sqrt{2\pi}R}^{L \sqrt{2\pi}R} \mathrm{d}X_0 =\sqrt{2\pi}R \vol(\mathbb{R})\ ,
\ee
where we imagined that the D-brane extends in some region $[-L,L]$. This again does not matter for the final result, we only need the factor $\sqrt{2\pi}R$ that arises from the correct normalization.

Finally, we note that the eigenvalues of the Laplacian $-\Delta$ are just $\ell(\ell+1)$. For Neumann boundary conditions, they have multiplicity $\ell+1$, whereas for Dirichlet boundary conditions, they have multiplicity $\ell$. Thus,
\begin{align}
\frac{T_{p+1}}{T_p}=\sqrt{2\pi}R \prod_{\ell=1}^\infty \sqrt{\frac{4\pi^2 \alpha' R^2}{\ell(\ell+1)}}=\frac{1}{\sqrt{2\pi\alpha'}} \prod_{\ell=1}^\infty \frac{1}{\sqrt{\ell(\ell+1)}}\ .
\end{align}
Since the result is independent of $R$, we made the convenient choice $R=\frac{1}{2\pi \sqrt{\alpha'}}$. The infinite product can be evaluated using zeta-function regularization.\footnote{Tree level partition functions in zeta-function regularization in string theory were considered in \cite{Grinstein:1986hd, Douglas:1986eu, Weisberger:1986qd}.} Define
\be 
\zeta_{\text{N}/\text{D}}(s)=\sum_{\ell=1}^\infty \frac{1}{(\ell(\ell+1))^s}\ .
\ee
We want to compute $\zeta_{\text{N}/\text{D}}'(0)$ which enters the regulated ratio of determinants. For this, we write
\begin{align}
\zeta_{\text{N}/\text{D}}(s)=\sum_{\ell=1}^\infty \left(\frac{1}{\ell^{2s}}-\frac{s}{\ell^{2s+1}}\right)+\sum_{\ell=1}^\infty \frac{1}{\ell^{2s}}\left(\frac{1}{(1+\ell^{-1})^s}-1+\frac{s}{\ell}\right)\ .
\end{align}
The first sum can be expressed through the Riemann zeta-function, whereas the second sum converges absolutely for $\Re s>-\frac{1}{2}$. Hence to evaluate the derivative at $s=0$, we can commute the derivative with the sum. We obtain
\be 
\zeta_{\text{N}/\text{D}}'(0)=2\zeta'(0)-\gamma+\sum_{\ell=1}^\infty \left(\frac{1}{\ell}-\log \left(1+\frac{1}{\ell}\right)\right)\ .
\ee
Here, we used already that the Riemann zeta-function behaves near $s=1$ as
\be 
\zeta(s)=\frac{1}{s-1}+\gamma+\mathcal{O}(s-1)\ ,
\ee
where $\gamma$ is the Euler-Mascheroni constant. Furthermore, we can use that $\zeta'(0)=-\frac{1}{2}\log(2\pi)$. The remaining sum is seen to be equal to $\gamma$ by definition:
\be 
\sum_{\ell=1}^n \left(\frac{1}{\ell}-\log \left(1+\frac{1}{\ell}\right)\right)=\sum_{\ell=1}^n \frac{1}{\ell}-\log(n+1) \overset{n \to \infty}{\longrightarrow} \gamma\ ,
\ee
where we used the the logarithmic piece is a telescoping sum. Finally, we simply obtain
\be 
\zeta_{\text{N}/\text{D}}'(0)=2\zeta'(0)=-\log(2\pi)\ .
\ee
Putting the pieces together gives
\be 
\frac{T_{p+1}}{T_p}=\frac{1}{\sqrt{2\pi\alpha'}} \exp\left(\frac{1}{2} \zeta_{\text{N}/\text{D}}'(0)\right)=\frac{1}{2\pi \sqrt{\alpha'}}\ . \label{eq:ratio tensions}
\ee
\subsection{Fixing normalization}\label{subsec:norm}
After having fixed the $p$-dependence, we can compute the overall normalization. We follow here the conventions of Polchinski \cite{Polchinski:1998rq}. We will compute the normalization for the D25-brane where we only impose Neumann boundary conditions. In his notation,
\be 
Z_\text{CFT}=C_{D_2}=\frac{1}{\alpha' g_\text{o}^2}\ ,
\ee
where $g_\text{o}$ is the open string coupling, compare to eq.~(6.4.14) in Polchinski.
We also have the following relation of the gravitational coupling $\kappa=\sqrt{8\pi G_\text{N}}$ to the open string coupling (eq.~(6.6.18) and eq.~(8.7.28)):
\be 
\kappa=2\pi g_\text{c}=2^{-17}\pi^{-\frac{23}{2}} (\alpha')^{-6} g_\text{o}^2 \ .
\ee
Finally, we should remember that the effective volume of the group $\mathrm{PSL}(2,\mathbb{R})$ is $-2\pi^2$ in Polchinski's normalization, see also the discussion in Appendix~\ref{app:volume PSL2R}. This is because the normalization of the ghosts lead to a different normalization of the measure on $\mathrm{PSL}(2,\mathbb{R})$ than the one we were considering above. 
Thus we can express the result for the D-brane tension as follows:
\be 
T_{25}=\frac{1}{2\pi^2} Z_\text{CFT}=\frac{1}{2\pi^2 \alpha' g_\text{o}^2}=\frac{\sqrt{\pi}}{16\kappa} (4\pi^2 \alpha')^{-7}\ .
\ee
For a general D$p$-brane, we combine this result with eq.~\!\eqref{eq:ratio tensions} and obtain
\be 
T_p=\frac{\sqrt{\pi}}{16\kappa} (4\pi^2 \alpha')^{\frac{11-p}{2}}\ .
\ee
This agrees with eq.~(8.7.26) of Polchinski and hence provides a simple way of computing D-brane tensions.

\section{Conclusions} \label{sec:conclusions}
We found that the disk partition function in string theory can be rigorously computed using standard path integral methods. Using one of the bosons on the worldsheet, one can further fix the residual gauge group $\mathrm{PSL}(2,\mathbb{R})$. We gave two possible gauge choices: in Section~\ref{sec:first gauge} we imposed that when expanding the boson $X$ into spherical harmonics, one of the coefficients is absent. In Section~\ref{sec:alternative gauge} we imposed that the derivative of $X$ vanishes at the origin of the worldsheet disk. Finally, in Section~\ref{sec:one point function} we used a more standard procedure and made use of the presence of a modulus in the worldsheet CFT which allows one to relate the result to a one-point function through conformal perturbation theory. In all these methods, the conclusion was the same: The group $\mathrm{PSL}(2,\mathbb{R})$ behaves as if it had a finite volume $-\frac{\pi^2}{2}$ in the path integral (for a suitable normalization of the metric on the group). We finally saw in Section~\ref{sec:D brane tension} that the disk partition function gives a very direct derivation of the D-brane tensions without the detours that are usually taken in the literature.
\medskip

In the following we mention some open questions and future directions.
\paragraph{Infinite volume.} We have given three independent computations of the disk partition function and to us they are quite convincingly showing that the gauge group $\mathrm{PSL}(2,\mathbb{R})$ should be thought of having finite volume. However, conceptually, this is somewhat counterintuitive. One starts in CFT with an integral over a function space $L^2(D)$ with Neumann or Dirichlet boundary conditions which is finite after an appropriate regularization. Gauging of $\mathrm{PSL}(2,\mathbb{R})$ identifies the gauge orbits, which are non-compact slices inside $L^2(D)$. If we would talk about a finite-dimensional integral, such an identification surely would lead to a vanishing result, due to the non-compactness of the gauge orbits. The finiteness of the result for the path integral is hence very unexpected and a result of an interesting interplay between the non-compactness of the gauge group and the subtleties of the path integral.

\paragraph{Sphere partition function.} Given our success with the disk partition function, one should ask whether one can similarly compute the more interesting sphere partition function in a similar manner. This does not seem to be the case from several perspectives.
\begin{enumerate}
\item Liu and Polchinski applied the same regularization procedure as for $\mathrm{PSL}(2,\mathbb{R})$ to the case of $\mathrm{PSL}(2,\mathbb{C})$. However, one also gets a logarithmic divergence in the cutoff that is akin to the appearance conformal anomaly in holographic renormalization \cite{Henningson:1998gx}. This prevents one from assigning a well-defined value to the volume.
\item The sphere partition function in flat space is expected to vanish. If we could perform a similar gauge fixing procedure as explored in this article using one flat spacetime direction, we would conclude that the sphere partition function should be vanishing for every background with a flat direction in it. This is not the case -- counterexamples include $c=1$ string theory and $\mathrm{AdS}_3 \times \mathrm{S}^3 \times \mathbb{T}^4$. Thus, one spacetime direction should not be sufficient to fix the gauge.
\item The sphere partition function should vanish for a compact target space. This is expected from supergravity where the on-shell action is a total derivative and hence vanishes for a compact spacetime. However, the ungauged worldsheet partition function is clearly non-vanishing and so $\mathrm{PSL}(2,\mathbb{C})$ needs to have an infinite volume for consistency.
\end{enumerate}
For these reasons, the computation of the sphere partition function is a much more subtle problem than the disk partition function that we have treated in this paper.

\section*{Acknowledgements}
We thank Raghu Mahajan for initial collaboration and very useful discussions. We also thank Adam Levine and Edward Witten for discussions and Douglas Stanford for comments on a preliminary draft of the paper. LE is supported by the IBM Einstein Fellowship at the Institute for Advanced Study. SP acknowledges the support from DOE grant DE-SC0009988.

\appendix
\section{Conventions} 
\subsection{The non-linear sigma-model}
We take the non-linear sigma-model on the worldsheet Riemann surface $\Sigma$ to be 
\be 
S[g,X]=\frac{1}{4\pi \alpha'}\int_\Sigma \mathrm{d}^2 z\ \sqrt{g} \, g^{ab} \partial_a X^\mu \partial_b X^\nu G_{\mu\nu}(X)\ ,
\ee
where $G_{\mu\nu}(X)$ is the spacetime metric. We will not have need of the $B$-field and the dilaton, since we assume throughout the text that there is one flat direction in spacetime that does not support a non-trivial $B$-field or a non-constant dilaton.

Let us review the gauge symmetries of the worldsheet action:
\begin{enumerate}
\item Diffeomorphism symmetry:
\begin{align}
X(z) &\longmapsto X \circ \varphi^{-1}(z)\ , \\
g_{ab}(z) &\longmapsto \frac{\mathrm{d}\varphi^c}{\mathrm{d}z^a}(\varphi^{-1}(z)) \frac{\mathrm{d}\varphi^d}{\mathrm{d} z^b}(\varphi^{-1}(z)) g_{cd}(\varphi^{-1}(z))\ .
\end{align}
for $\varphi:\Sigma \longmapsto \Sigma$ a diffeomorphism.
\item Weyl symmetry:
\begin{align}
g_{ab}(z) &\longmapsto \lambda(z) g_{ab}(z)
\end{align}
for some positive function $\lambda:\Sigma \longmapsto \mathbb{R}_{>0}$.
\end{enumerate}
Conformal gauge fixes $g=\hat{g}$ for some reference metric $\hat{g}$ on $\Sigma$. In the case of $\Sigma=\mathrm{S}^2$ or $\Sigma=D$ in the open string case, this gauge is always attainable. For example, in Section~\ref{sec:first gauge} we have considered $D$ and $\hat{g}$ is given by eq.~\eqref{eq:metric}. For higher genus surfaces there would be a moduli space of inequivalent metrics which is the moduli space of Riemann surfaces. It is well-known that the Weyl symmetry is anomalous unless we are considering the critical string. We will assume throughout the text that the string is critical.
\subsection{Spherical harmonics on the disk}\label{app:spherical harmonics}
In this Appendix, we fix our conventions for spherical harmonics. They take the following form on the unit disk parametrized by the complex coordinates $(z,\bar{z})$:
\begin{multline}
Y_{\ell,m}(z,\bar z)=\sqrt{\frac{(2 \ell+1) (\left| m\right| +\ell)!}{2 \pi  (\ell-\left| m\right| )!(|m|!)^2}} \, (z \bar z+1)^{\ell+1} \\
\times {} _2F_1(\ell+1,\ell+\left| m\right| +1;\left| m\right| +1;-z\bar z)\begin{cases}z^m \quad \ m\geq 0\\
\bar z^m\ \quad\ m\leq 0\ .
\end{cases}
\end{multline}
Spherical harmonics satisfy
\begin{subequations}
\begin{align}
\text{Dirichlet boundary conditions for }\ell+m &\in 2\mathbb{Z}+1\ , \\
\text{Neumann boundary conditions for }\ell+m &\in 2\mathbb{Z}\ .
\end{align}
\end{subequations}
They are orthonormal on the disk with round metric \eqref{eq:metric} (and hence differ by the usual normalization of  spherical harmonics by a factor of $\sqrt{2}$, since those are orthornomal on the sphere)
\begin{equation}
\int_{D} \frac{4r\,  \mathrm{d}r \, \mathrm{d}\theta}{(1+r^2)^2} Y_{\ell,m}(r e^{i\theta},r e^{-i\theta})Y_{\ell',m'}(r e^{-i\theta},r e^{i\theta})= \delta_{\ell\ell'}\delta_{mm'}\ ,
\end{equation}
where both $(\ell,m)$ and $(\ell',m')$ satisfy the same boundary condition. 
Spherical harmonics are eigenfunctions of the Laplacian on the disk with the upper hemisphere metric,
\begin{equation}
\Delta Y_{\ell,m}=\frac{1+r^2}{4}\left(r^{-1} \partial_r \left(r \partial_r Y_{\ell,m}\right) +r^{-2}\partial_\phi^2 Y_{\ell,m}\right)=-\ell(\ell+1)Y_{\ell,m}\ .
\end{equation}

\section{The regularized volume of \texorpdfstring{$\boldsymbol{\mathrm{PSL}(2,\mathbb{R})}$}{PSL(2,R)}}\label{app:volume PSL2R}
In this Appendix, we review the computation of the regularized volume of the M\"obius group $\mathrm{PSL}(2,\mathbb{R})$ following \cite{Liu:1987nz}.

The group of M\"{o}bius transformations preserving the unit disk is
\be 
\mathrm{PSL}(2,\mathbb{R}) \cong \mathrm{PSU}(1,1)=\left\{ \begin{pmatrix}
a & b \\ \bar{b} & \bar{a} 
\end{pmatrix} \, \Big| \, |a|^2-|b|^2=1 \right\}\Big/ \sim\ .
\ee
Here, the equivalence $\sim$ identifies the matrix with the negative matrix. Let us parametrize 
\be
a=\mathrm{e}^{i\phi}\cosh x\,,\qquad b=\mathrm{e}^{i\psi}\sinh x\,, \qquad x\in[0,\infty)\,, \quad\phi\in[0,2\pi)\,,\quad \psi\in [0,\pi)\ .
\ee
The range of $\psi$ indicates that we are dealing with $\mathrm{PSL}(2,\mathbb{R})$ rather than the usual $\mathrm{SL}(2,\mathbb{R})$ in which case $\psi$ would have run from $0$ to $2\pi$. The formal volume of the group is given by

\be
\mathrm{vol}\left(\mathrm{PSL}(2,
\mathbb{R})\right)=\frac{1}{2}\int\ \mathrm{d}^2a\ \mathrm{d}^2b\ \delta\left(|a|^2-|b|^2-1\right)\ .
\ee
In terms of $(x,\phi,\psi)$, the formal expression for the  volume of  $\mathrm{PSL}(2,
\mathbb{R})$ becomes
\be
\int_{0}^{\infty} \mathrm{d}x \int_{0}^{2\pi} \mathrm{d}\phi\int_{0}^{\pi}\mathrm{d}\psi\ \cosh x \sinh x\ .
\ee
Of course the above is divergent, so we need to regulate it. The prescription advocated by Polchinski and Liu \cite{Liu:1987nz} is to cut off the $x$ integral at some large radius $x=x_*$, which leads us to the following expression
\be\label{volumeWc}
\mathrm{vol}\left(\mathrm{PSL}(2,\mathbb{R})\right)=2\pi^2 \int_{0}^{x_*}\mathrm{d}x\ \cosh x \sinh x=\pi^2 \sinh^2 x_*\ .
\ee
The area of the cutoff surface at $x=x_*$ is equal to
\be
A_*=2\pi^2\sinh x_* \cosh x_*\ .
\ee
Thus we have
\be
\mathrm{vol}\left(\mathrm{PSL}(2,\mathbb{R})\right)=\frac{1}{2}\left[\sqrt{\pi^4+A_*^2}-\pi^2\right]\underset{A_*\to \infty}{\simeq} \frac{A_*}{2}-\frac{\pi ^2}{2}+O\left(A_*^{-1}\right) \ .
\ee
To obtain a finite answer for the volume one proceeds as in the gravitational path integral and adds a local counter term on the cutoff surface. Thus, the regularized volume is defined as
\be 
\mathrm{vol}\left(\mathrm{PSL}(2,\mathbb{R})\right)_{\mathrm{reg}}=\lim_{x_* \to \infty} \int_{\mathrm{G}_*} \mathrm{d}^3 x\ \sqrt{g} -\frac{1}{2} \int_{\partial \mathrm{G}_*} \mathrm{d}^2 x\ \sqrt{h}\ ,
\ee
where $\mathrm{G}_*$ is the group manifold with a cutoff at $x_*$ and $h$ is the induced metric on the cutoff surface. In the case of $\mathrm{PSL}(2,\mathbb{R}) \cong \mathrm{PSU}(1,1)$, this leads to
\begin{tcolorbox}
\be
\mathrm{vol}\left(\mathrm{PSL}(2,\mathbb{R})\right)_{\mathrm{reg}}=-\frac{\pi^2}{2}\ ,\quad \text{\`a la Liu-Polchinski  \cite{Liu:1987nz}.}
\ee
\end{tcolorbox}
Several comments are in order.
\begin{enumerate}
\item This result is independent of how exactly we choose the cutoff surface. 
\item Since $\mathrm{PSL}(2,\mathbb{R})/\mathrm{U}(1) \cong \text{Euclidean }\mathrm{AdS}_2$, this computation is exactly analogous (after integrating out $\phi$) to the computation of the gravitational on-shell action in $\mathrm{AdS}_2$.
\item This result depends of course on the normalization of the metric on $\mathrm{PSL}(2,\mathbb{R})$. We have chosen a normalization such that $\mathrm{PSU}(1,1)$ is realized as a quadric in $\mathbb{C}^2 \cong \mathbb{R}^4$ with unit radius. Equivalently our normalization is fixed by requiring that the Ricci scalar is $\mathcal{R}=-6$ on the group manifold. This is not the normalization that is often employed in string theory. Instead one parametrizes a group element $\mathrm{PSL}(2,\mathbb{R})$ by the three images of $0$ and $1$ and $\infty$: $\gamma(0)=x_1$, $\gamma(1)=x_2$ and $\gamma(\infty)=x_3$ and takes the measure to be the one of the ghost 3-point function in the standard normalization,
\be 
\mathrm{d}\mu = \frac{\mathrm{d}x_1 \, \mathrm{d}x_2 \, \mathrm{d}x_3}{|(x_1-x_2)(x_2-x_3)(x_3-x_1)|}\ . \label{eq:ghost 3-point function measure}
\ee
However, by relating the measure in the $(x,\psi,\phi)$ variables that we considered above to the coordinates $(x_1,x_2,x_3)$, one finds that the two measures differ by a factor of 4. The relevant change of variables is somewhat lengthy, but for example the change of variables $\theta_i =2\arctan x_i$ transforms this measure to the canonical measure on the unit disc that is also discussed in \cite[eq.~(7)]{Liu:1987nz}.
In the measure that is defined by the ghosts via eq.~\eqref{eq:ghost 3-point function measure}, the regularized volume of $\mathrm{PSL}(2,\mathbb{R})$ instead works out to be $4 \times (-\frac{\pi^2}{2})=-2\pi^2$.
\item If one repeats the same computation for $\mathrm{PSL}(2,\mathbb{C})$ (which is the relevant group for the sphere partition function) one finds an obstruction. The reason is well known: this computation is essentially the same as computing the on-shell action of gravity on Euclidean $\mathrm{AdS}_3 \cong \mathrm{PSL}(2,\mathbb{C})/\mathrm{SU}(2)$, which suffers from the conformal anomaly. The conformal anomaly leads to a term that is logarithmically divergent in the cutoff and which cannot be removed by any local counterterm. Thus, one cannot give a sensible value for the volume of $\mathrm{PSL}(2,\mathbb{C})$.
\end{enumerate}

\section{The ``Signed'' Faddeev-Popov procedure} \label{app:FP procedure}
Let us review the Faddeev-Popov procedure. The gauges that we have chosen involve Gribov copies and we have to be careful to deal with them correctly. This means that while the gauge is admissible, it usually is not uniquely so \cite{Gribov:1977wm}.
Because of them we will use a slightly different version of the FP-procedure that counts intersections of gauge orbits with the gauge slight with a sign according to their intersection number. The procedure we will use was proposed in \cite{Hirschfeld:1978yq} as a solution to the problem of Gribov copies. 

Let $\mathcal{G}$ be the gauge group in question, which in our case is $\mathrm{PSL}(2,\mathbb{R})$. We want to compute
\be 
Z=\int \frac{\mathscr{D}X}{\mathop{\text{vol}}(\mathcal{G})} \mathrm{e}^{-S[X]}\ ,
\ee
where the domain of the path integral is given by the appropriate function space.
We write the action of the gauge group on the fields $X$ as $g \cdot X\equiv X^g$. We assume that the gauge group and the measure are invariant under the gauge group (that is, the gauge symmetry is non-anomalous). So $S[X^g]=S[X]$ and $\mathscr{D}X^g=\mathscr{D}X$. The latter assumption requires of course again the inclusion of the other matter fields and the ghosts on the worldsheet which we tacitly assume to be included in the calculation. We also assume for simplicity that there are no large gauge transformations, i.e.~$\mathcal{G}$ is connected. This is the case in our example. 

One then starts by inserting the identity
\be 
1= \int_\mathcal{G} \mathrm{d}g\ \Delta(X^g)\delta(F(X^g)) \label{eq:def Delta}
\ee
in the path integral. Here, $F(X)$ is the gauge fix condition that (ideally) picks one representative of every gauge orbit. There are subtleties when this is not the case. 

\begin{figure}[!ht]
\begin{center}
\includegraphics[width=.5\textwidth]{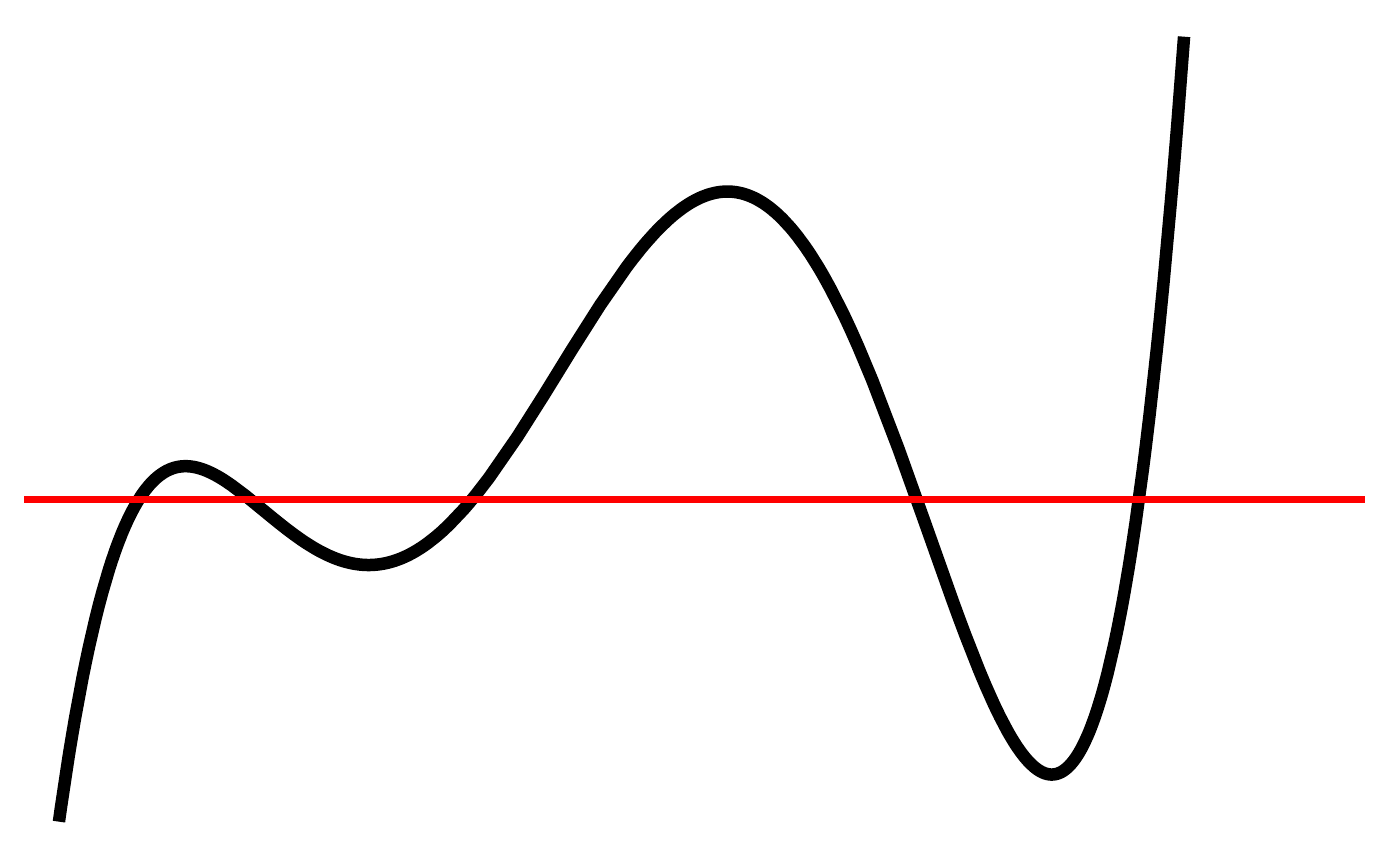}
\end{center}
\caption{The intersection number is $1$ while the total number of roots are $5$. The horizontal red line is $y=0$, the gauge fixing line while the black curve is the function $f$. The gauge choice is $f(x)=0$ which provides $5$ roots. The contribution of them towards signed intersection number is $1$.} \label{fig:gauge}
\end{figure}

For illustration\footnote{We thank Dalimil Maz\'{a}\v{c} for pointing us to a lecture by Davide Gaiotto where similar toy example is considered \cite{Gaiotto_lecture}.}, let us consider the gauge group $\mathbb{R}$ with a gauge constraint, which is implemented by the function $f(x)=0$. The analogous identity reads
\be 
1=\int_{-\infty}^{\infty}\mathrm{d}x\  |f'(x)| \delta(f(x)) 
\ee
if the $f(x)=0$ has only one solution at $x=x_*$. This is the situation where the gauge condition picks a unique representative in the gauge orbit. If $f(x)=0$ has multiple solutions, we have instead
\begin{equation}
\int_{-\infty}^{\infty}\mathrm{d}x\ |f'(x)| \delta(f(x))=\sum_{x:\, f(x)=0}1=\ \text{number of roots of}\ f\ .
\end{equation}
So we cannot directly insert this in the path integral and expect a simple answer. Instead, we have to restrict the integral to a region where $f(x)=0$ has only solution. This is the usual Gribov problem. Nonetheless one can bypass this problem if one assume suitable boundary conditions on the function $f$. For example, assume that $f$ has to following additional property
\begin{equation}\label{eq:boundaryc}
\lim_{x\to\pm\infty} f(x) = \pm \infty\ .
\end{equation}
In this case, we have the identity
\begin{equation}\label{eq:WC}
\int_{-\infty}^{\infty} \mathrm{d}x\ f'(x) \delta(f(x))= \sum_{x: \, f(x)=0} \mathrm{sgn}\left(f'(x)\right)=1\ ,
\end{equation}
where the last equality follows from the boundary condition eq.~\!\eqref{eq:boundaryc}. In fact $1$ is the intersection number of the graph $y=f(x)$ with $y=0$ in the sense of topology where intersections are counted with signs. See Figure~\ref{fig:gauge} for an illustration.
In this toy set up, omission of the absolute value of $f'(x)$ removes the Gribov ambiguities. 
In what follows we will be using the above kind of signed FP procedure, but it will involve more than one variable. Furthermore, we are required to justify that the intersection number is an invariant among the space of functions that we are dealing with while doing the path integral. This is not obvious since the gauge group is non-compact and there might be similar boundary conditions. The corresponding identity is
\begin{equation}\label{eq:Gauge}
\int_\mathcal{G} \mathrm{d}g\  \mathop{\text{det}} \mathop{\text{Jac}} F(X^g) \, \delta(F(X^g))=\sum_{g:\, F(X^g)=0} \mathop{\text{sgn}}\left(\mathop{\text{det}} \mathop{\text{Jac}} F(X^g)\right)=\mathcal{I}\ .
\end{equation}
The RHS is in fact an intersection number, which has a chance to be independent of $X$, so that the LHS can be inserted in the path integral. Let us assume this for now, we will justify below that this is indeed the case for the situation of interest.

Let us now insert 
\begin{equation}\label{eq:id}
1
=\int_\mathcal{G} \mathrm{d}g\ \Delta(X^g) \delta(F(X^g))\,,\quad\ \Delta(X)\equiv \frac{1}{\mathcal{I}} \mathop{\text{det}} \mathop{\text{Jac}} F(X)
\end{equation}
 in the path integral to obtain

\begin{align}
Z&=\int \frac{\mathscr{D}X}{\mathop{\text{vol}}(\mathcal{G})} \int_\mathcal{G} \mathrm{d}g\ \Delta(X^g) \delta(F(X^g))\mathrm{e}^{-S[X]}\\
&= \int_\mathcal{G} \mathrm{d}g \int \frac{\mathscr{D}X^g}{\mathop{\text{vol}}(\mathcal{G})} \Delta(X^g) \delta(F(X^g))\mathrm{e}^{-S[X^g]} \\
&= \int_\mathcal{G} \mathrm{d}g \int \frac{\mathscr{D}X}{\mathop{\text{vol}}(\mathcal{G})} \Delta(X)  \delta(F(X))\mathrm{e}^{-S[X]}\ .
\end{align}
In the second line we used the invariance of various quantities under the group action. In the third line we replaced the dummy variable $X^g$ with $X$ everywhere. Now nothing depends on $g$ anymore and we can formally cancel $\mathop{\text{vol}}(\mathcal{G})$ with $\int_\mathcal{G} \mathrm{d}g$. One hence obtains
\be 
Z=\int\mathscr{D}X\ \Delta(X)  \delta(F(X))\mathrm{e}^{-S[X]}\ .
\ee
The only difference to the standard Faddeev-Popov procedure is a missing absolute value sign for $\Delta(X)=\frac{1}{\mathcal{I}} \mathop{\text{det}} \mathop{\text{Jac}} F(X)$.

\bibliographystyle{JHEP}
\bibliography{bib}

\providecommand{\href}[2]{#2}\begingroup\raggedright\begin{thebibliography}{10}

\bibitem{DHoker:1988pdl}
E.~D'Hoker and D.~Phong, \emph{{The Geometry of String Perturbation Theory}},
  \href{https://doi.org/10.1103/RevModPhys.60.917}{\emph{Rev. Mod. Phys.}
  {\bfseries 60} (1988) 917}.

\bibitem{Witten:2012bh}
E.~Witten, \emph{{Superstring Perturbation Theory Revisited}},
  \href{https://arxiv.org/abs/1209.5461}{{\ttfamily 1209.5461}}.

\bibitem{Tseytlin:1987ww}
A.~A. Tseytlin, \emph{{Renormalization of Mobius Infinities and Partition
  Function Representation for String Theory Effective Action}},
  \href{https://doi.org/10.1016/0370-2693(88)90857-X}{\emph{Phys. Lett. B}
  {\bfseries 202} (1988) 81}.

\bibitem{Liu:1987nz}
J.~Liu and J.~Polchinski, \emph{{Renormalization of the Mobius Volume}},
  \href{https://doi.org/10.1016/0370-2693(88)91566-3}{\emph{Phys. Lett. B}
  {\bfseries 203} (1988) 39}.

\bibitem{Tseytlin:1988tv}
A.~A. Tseytlin, \emph{{Mobius Infinity Subtraction and Effective Action in
  $\sigma$ Model Approach to Closed String Theory}},
  \href{https://doi.org/10.1016/0370-2693(88)90421-2}{\emph{Phys. Lett. B}
  {\bfseries 208} (1988) 221}.

\bibitem{Erbin:2019uiz}
H.~Erbin, J.~Maldacena and D.~Skliros, \emph{{Two-Point String Amplitudes}},
  \href{https://doi.org/10.1007/JHEP07(2019)139}{\emph{JHEP} {\bfseries 07}
  (2019) 139} [\href{https://arxiv.org/abs/1906.06051}{{\ttfamily
  1906.06051}}].

\bibitem{Maldacena:2001km}
J.~M. Maldacena and H.~Ooguri, \emph{{Strings in ${\rm AdS}_3$ and ${\rm
  SL}(2,\mathds{R})$ WZW model. Part 3. Correlation functions}},
  \href{https://doi.org/10.1103/PhysRevD.65.106006}{\emph{Phys. Rev.}
  {\bfseries D65} (2002) 106006}
  [\href{https://arxiv.org/abs/hep-th/0111180}{{\ttfamily hep-th/0111180}}].

\bibitem{Troost:2011ud}
J.~Troost, \emph{{The $AdS_3$ central charge in string theory}},
  \href{https://doi.org/10.1016/j.physletb.2011.10.007}{\emph{Phys. Lett. B}
  {\bfseries 705} (2011) 260}
  [\href{https://arxiv.org/abs/1109.1923}{{\ttfamily 1109.1923}}].

\bibitem{Gibbons:1976ue}
G.~W. Gibbons and S.~W. Hawking, \emph{{Action Integrals and Partition
  Functions in Quantum Gravity}},
  \href{https://doi.org/10.1103/PhysRevD.15.2752}{\emph{Phys. Rev. D}
  {\bfseries 15} (1977) 2752}.

\bibitem{Polchinski:1995mt}
J.~Polchinski, \emph{{Dirichlet Branes and Ramond-Ramond Charges}},
  \href{https://doi.org/10.1103/PhysRevLett.75.4724}{\emph{Phys. Rev. Lett.}
  {\bfseries 75} (1995) 4724}
  [\href{https://arxiv.org/abs/hep-th/9510017}{{\ttfamily hep-th/9510017}}].

\bibitem{Shapiro:1975cz}
J.~A. Shapiro, \emph{{On the Renormalization of Dual Models}},
  \href{https://doi.org/10.1103/PhysRevD.11.2937}{\emph{Phys. Rev. D}
  {\bfseries 11} (1975) 2937}.

\bibitem{Ademollo:1975pf}
M.~Ademollo, A.~D'Adda, R.~D'Auria, F.~Gliozzi, E.~Napolitano, S.~Sciuto
  et~al., \emph{{Soft Dilations and Scale Renormalization in Dual Theories}},
  \href{https://doi.org/10.1016/0550-3213(75)90491-5}{\emph{Nucl. Phys. B}
  {\bfseries 94} (1975) 221}.

\bibitem{Hawking:1976ja}
S.~W. Hawking, \emph{{Zeta Function Regularization of Path Integrals in Curved
  Space-Time}}, \href{https://doi.org/10.1007/BF01626516}{\emph{Commun. Math.
  Phys.} {\bfseries 55} (1977) 133}.

\bibitem{Grinstein:1986hd}
B.~Grinstein and M.~B. Wise, \emph{{Vacuum Energy and Dilaton Tadpole for the
  Unoriented Closed Bosonic String}},
  \href{https://doi.org/10.1103/PhysRevD.35.3285}{\emph{Phys. Rev. D}
  {\bfseries 35} (1987) 655}.

\bibitem{Douglas:1986eu}
M.~R. Douglas and B.~Grinstein, \emph{{Dilaton Tadpole for the Open Bosonic
  String}}, \href{https://doi.org/10.1016/0370-2693(87)91416-X}{\emph{Phys.
  Lett. B} {\bfseries 183} (1987) 52}.

\bibitem{Weisberger:1986qd}
W.~I. Weisberger, \emph{{Normalization of the Path Integral Measure and the
  Coupling Constants for Bosonic Strings}},
  \href{https://doi.org/10.1016/0550-3213(87)90032-0}{\emph{Nucl. Phys. B}
  {\bfseries 284} (1987) 171}.

\bibitem{Polchinski:1998rq}
J.~Polchinski, \emph{{String theory. Vol. 1: An introduction to the bosonic
  string}}, Cambridge Monographs on Mathematical Physics. Cambridge University
  Press, 12, 2007,
  \href{https://doi.org/10.1017/CBO9780511816079}{10.1017/CBO9780511816079}.

\bibitem{Henningson:1998gx}
M.~Henningson and K.~Skenderis, \emph{{The Holographic Weyl Anomaly}},
  \href{https://doi.org/10.1088/1126-6708/1998/07/023}{\emph{JHEP} {\bfseries
  07} (1998) 023} [\href{https://arxiv.org/abs/hep-th/9806087}{{\ttfamily
  hep-th/9806087}}].

\bibitem{Gribov:1977wm}
V.~N. Gribov, \emph{{Quantization of Nonabelian Gauge Theories}},
  \href{https://doi.org/10.1016/0550-3213(78)90175-X}{\emph{Nucl. Phys. B}
  {\bfseries 139} (1978) 1}.

\bibitem{Hirschfeld:1978yq}
P.~Hirschfeld, \emph{{Strong Evidence That Gribov Copying Does Not Affect Gauge
  Theory Functional Integral}},
  \href{https://doi.org/10.1016/0550-3213(79)90052-X}{\emph{Nucl. Phys. B}
  {\bfseries 157} (1979) 37}.

\bibitem{Gaiotto_lecture}
D.~Gaiotto, ``String theory.''
  \url{http://pirsa.org/displayFlash.php?id=15010066}, 2015.

\end{thebibliography}\endgroup
\end{document}